\documentclass[conference]{IEEEtran}
\usepackage{graphicx}
\usepackage{subfigure}
\usepackage{float}
\usepackage{algorithm}
\usepackage{algorithmic}
\usepackage{mathrsfs}
\usepackage{amsfonts}
\usepackage{array}
\usepackage{amssymb,amsmath}
\usepackage{longtable}
\usepackage{rotating}
\usepackage{multirow}
\usepackage{epstopdf}
\usepackage{cite}
\usepackage{float}
\usepackage[marginal]{footmisc}
\usepackage{stfloats}
\usepackage[colorlinks,
            linkcolor=black,
            anchorcolor=black,
            urlcolor=blue,
            citecolor=black]{hyperref}
\usepackage[hyphenbreaks]{breakurl}
\usepackage{geometry}
\renewcommand{\baselinestretch}{1}

\def\BibTeX{{\rm B\kern-.05em{\sc i\kern-.025em b}\kern-.08em
    T\kern-.1667em\lower.7ex\hbox{E}\kern-.125emX}}

\renewcommand{\baselinestretch}{1}
\begin{document}

\title{Multi-Resource Allocation for On-Device Distributed Federated Learning Systems}
\author{\IEEEauthorblockN{Yulan Gao\textsuperscript{1,2} Ziqiang Ye\textsuperscript{1}  Han Yu\textsuperscript{2} Zehui Xiong\textsuperscript{3} Yue Xiao\textsuperscript{1} Dusit Niyato\textsuperscript{2} }

\IEEEauthorblockA{\textsuperscript{1}{The National Key Laboratory of Science and Technology on Communications} \\
{University of Electronic Science and Technology of China}\\
{email: xiaoyue@uestc.edu.cn}\\
}

\IEEEauthorblockA{\textsuperscript{2}{The School of Computer Science and Engineering,}
{Nanyang Technological University, Singapore 639798}\\
{email: han.yu@ntu.edu.sg}
}

\IEEEauthorblockA{\textsuperscript{3}{Pillar of Information Systems Technology and Design}\\
{Singapore University of Technology and Design (SUTD), Singapore 487372} \\
}
}

\maketitle

\begin{abstract}
This work poses a distributed multi-resource allocation scheme for minimizing the weighted sum of latency and energy consumption in the on-device distributed federated learning (FL) system.
Each mobile device in the system engages the model training process within the specified area and allocates its computation and communication resources for deriving and uploading parameters, respectively, to minimize the objective of system subject to the computation/communication budget and a target latency requirement.
In particular, mobile devices are connect via wireless TCP/IP architectures.
Exploiting the optimization problem structure, the problem can be decomposed to two convex sub-problems.
Drawing on the Lagrangian dual and harmony search techniques, we characterize the global optimal solution by the closed-form solutions to all sub-problems, which give qualitative insights to multi-resource tradeoff.
Numerical simulations are used to validate the analysis and assess the performance of the proposed algorithm.

\end{abstract}
\IEEEpeerreviewmaketitle

\section{Introduction}

\IEEEPARstart{O}{wing} to the ever-growing volume of data traffic and a pervasive introduction of artificial intelligence tools, such as machine learning (ML) particularly deep learning, we are on the edge of evolution \cite{merluzzi2021wireless}.
Nowadays the traditional ML is powered by cloud-centric approach that relies on a cloud-based server or data center with the broad accessibility of computation, storage and the whole dataset.
However, the prolific spread of intelligent mobile devices and the ever-growing high-stake applications, the biggest challenge in these context is to meet the soaring demand for computation/communication resource required to ensure latency sensitive computation and privacy.
Naturally, the traditional cloud-centric ML methodologies are no longer sustainable \cite{zhu2020toward}.
Meanwhile, with the increasingly advanced sensors, computing, and communication capabilities equipped to intelligent mobile devices,  performing training directly at the edge device is a promising way out of this gridlock, which is commonly referred as edge ML \cite{park2019wireless}.
However, by nature, most mobile devices are in general resource-constrained, e.g., limited computation, storage, and battery, etc.
Therefore, a significant challenge would be posed by this intradevice conflict for the future edge ML platform development.
A burgeoning ML approach called federated learning (FL) has been introduced in \cite{mcmahan2016federated} makes it is  possible to facilitate collaborative ML among distributed devices and enjoy the benefits of better privacy and less communication resources.

To realize privacy, low latency, and highly efficient use of network bandwidth, FL is an enabling technology for ML model training at mobile edge networks.
Due to its promising potential, a lot of research attention is focused on enabling low-latency and energy-efficient resource management in the area of on-device distributed FL \cite{wang2018edge}.
More precisely, several advanced optimization algorithms have been used to speed up the training process by taking advantages of computing power and distributed data over multiple devices.
As mentioned in \cite{tran2019federated}, in the FL context, it is important to control the reliability in terms of respecting the perspective of communication and computation, evaluating the accuracy of the decisions taken by the edge server, which can involve fulfilling tasks such as prediction, estimation, classification, etc.
In a nutshell, the goal of FL is to devise resource allocation strategies that enable ML at the wireless network edge with low energy consumption, low E2E latency and high learning/inference accuracy.

In this paper, we consider a multi-resource allocation problem for on-device distributed FL.
The wireless TCP/IP protocol is adopted, which  describes the steady state of connective among mobile devices.
Our main contribution in this paper is to minimize the weighted sum of latency and energy consumption by jointly optimizing communication and computation resources to meet the system constraints while guaranteeing a prescribed performance of training model.
Specifically, using Lagrangian dual theory and harmony search, we develop a low complexity and provably convergent optimization algorithm to tackle the Min-Max problem that account for the latency constraint of mobile devices.
Exploiting the framework of alternating minimization, the global optimal solution can be characterized by the closed-form solutions to sub-problems, which given qualitative insights to diverse resources tradeoff. 
Additionally, from the simulation results, we can conclude that the proposed alternating algorithm effectively optimizes devices' energy consumption while guaranteeing low latency.

The remainder of this paper is organized as follows.
In Section II, we introduce the system model and problem formulation which includes
the connected distributed mobile devices modeling,  on-device distributed federated learning, communication model, computation model, and problem formulation.
In Section III, we propose the scheduling algorithm design.
The simulation results and conclusions are shown in Section IV and V, respectively.

\section{System Model and Problem Formulation }

\begin{figure}[!t]
	\centering
	\includegraphics[width=2.75in]{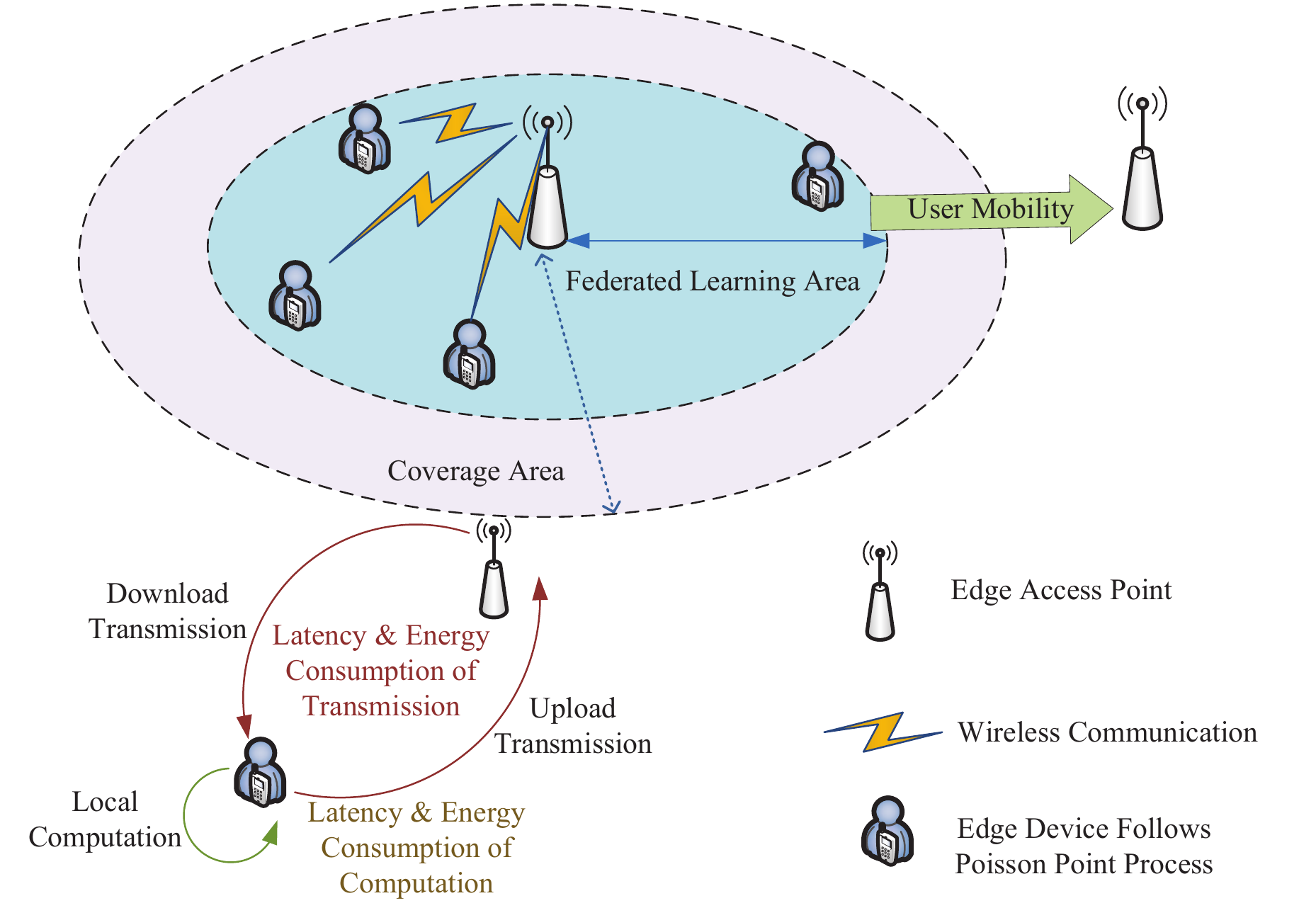}
	\caption{System model.}
	\label{fig:1}
\end{figure}

Consider an on-device distributed FL system in the urban scenario that consists of $N$ mobile devices and there are many edge servers (e.g., small base stations, access points) distributed over the service area, which are connected by fiber links.
As shown in Fig. \ref{fig:1}, mobile devices are only eligible participants in the current FL  after entering the specified area smaller than the coverage area of edge server, called the FL area.
Assuming mobile devices covered by an edge server,  each mobile device in set ${\cal N}$ realizes the communication with the edge server through the rapidly evolving C-V2X technology proposed by {\sl Apostolos Company and Intel Corporation} \cite{papathanassiou2017cellular}.
In addition, as aforementioned the edge servers are linked through fiber, and thus we omit the communication latency between them.

\subsection{On-Device Distributed FL System}
On-device distributed FL system allows mobile devices to collaboratively compute a shared model while keeping personal data remains local, which enjoys several benefits such as low-latency, low-power consumption as well as alleviating their privacy concerns.
In typical FL problem, each mobile device $n\in{\cal N}$ has a private dataset ${\cal D}_n$.
The task of each data owner $n$ is to find the model parameter ${\boldsymbol w}\in {\mathbb R}^d$ to minimize $f({\boldsymbol w})=1/D\sum_{n=1}^N\sum_{i\in{\cal D}_n} f_i({\boldsymbol w}),$  where $D=\sum_{n=1}^ND_n$ is the total number of data points and $D_n$ is the size of dataset ${\cal D}_n,$ and $f_i({\boldsymbol w}), 1\leq i\leq  D_n, \forall  n\in{\cal N}$ is defined as the loss function that characterizes the output $y_i$ for a sample data $\{{\boldsymbol x}_i, y_i\}, \forall 1\leq i\leq D_n.$

To compute the optimal parameter ${\boldsymbol w}$,  each mobile device iteratively computes the local model ${\boldsymbol w}_n$ and gradient until a local accuracy $0\leq \theta\leq 1$ is achieved and upload them to the edge server.
Then, all collected local parameters and gradients are aggregated to generate a global model $\boldsymbol w$.
When a specific global accuracy $0\leq \epsilon\leq 1$ is reached, the entire training process is terminated.
Upon achieving an global accuracy level $\epsilon$, a number of global iterations are required, causing interaction between mobile devices and edge server.
As mentioned in \cite{tran2019federated}, the upper bound of global iterations is closely related to the local accuracy $\theta$ and the global accuracy $\epsilon,$  which is specifically expressed as
$K(\epsilon, \theta)={\cal O}(\log(1/\epsilon))/(1-\theta)$.
The upper bound of local computation iterations is normalized to $\log(1/\theta)$.
Likewise, for a fixed global accuracy $\epsilon$, so $K(\epsilon, \theta)$ can also be normalized to $K(\theta)=1/(1-\theta)$.

\subsection{Communication Model}

The interaction between mobile device $n$ and the edge server is based on the wireless TCP/IP protocol.
In the spirit of the work presented in \cite{baccarelli2003optimized} , the packets-delays can be modeled as an i.i.d random sequence $\{\Delta_{\text{IP}}(t)\in \mathbb{R}_0^+,t\ge 1 \}$.
The probability density function of  $\Delta_{\text{IP}}$ follows a uniform distribution over interval $[0,\Delta_{\text{IP}}^{\max}]$, where $\Delta_{\text{IP}}^{\max}$ is the maximum packet delay of the IP layer which can be measured in multiple time slots.
Assuming the value of $\Delta_{\text{IP}}^{\max}$ is known to facilitate problem analysis.
Accordingly, the Round Trip Time (RTT) can be calculated by iteration with the following method:
\begin{equation}\label{eq:5}
\begin{aligned}
\texttt{RTT}(t)&=0.75\texttt{RTT}(t-1)+0.25\Delta_{\text{IP}}(t), ~t\ge 1,\\
 \texttt{RTT}(0)&=0.
\end{aligned}
\end{equation}

Following \cite{gudmundson1991correlation},   the data rate of mobile device $n\in{\cal N}$  can thus be written as
\begin{equation}\label{eq:6}
r_n=\sigma_n(t)\sqrt[]{p_n},
\end{equation}
where $p_n$ is the transmit power of device $n$ and $\sigma_n = {K_0\sqrt[]{Z_n}}/{\texttt{RTT}[t]}, t\geq 1$ represents the connection state at time slot $t$.
Taking mobility into account, $Z(t)$ is modeled as a time-correlated and log-distributed sequence $\{Z(t) \in \mathcal R_0^+,t\ge 1\}$, i.e.,  $Z(t) \triangleq a_0 10^{0.1x(t)}, \forall t\ge 1$, where $a_0 \approx 0.9738$, the sequence $\{x(t),t\ge 1\}$ is a time-dependent, zero-mean, unit-variance and stationary Markov random sequence with the probability density function uniformly distributed in the interval $[-\sqrt[]{3},~\sqrt[]{3}]$.
Moreover, $K_0$  is commonly regarded as a positive constant with the following formula:
\begin{equation}\label{eq:8}
K_{0}\triangleq((3/2b)^{1/2} \texttt{MSS})/(C+(A/CB^{2}) \Gamma(1 ; CB))^{1 / 2},
\end{equation}
where $b=2$, $\texttt{MSS}$ (byte) is the {\sl Maximum Segment Size} which can be defined as the maximum permitted size of the segment.
$\Gamma (\cdot ,\cdot )$ is the Gamma function.
The positive constants $A, B,$ and $C$ represent the performance of error in the {\sl Forward Error Correction} system.

We let $C_n$ denote the size of local parameters in mobile device $n$, so the energy consumption for uploading parameters at device $n$ takes the form $\text{E}_n^{\text{up}}=\frac{C_n\sqrt[]{p_n}}{\sigma_n},$ where the corresponding uploading time is $\text{T}_n^{\text{up}}=C_n/r_n.$

\subsection{Computation Model}

We use a tuple $J_n\triangleq\{ D_n, B_n\}$ to represent the computationally task of mobile device $n$ training the local model, in which $B_n$ is the number of CPU cycles that is required to process a data sample.
Let $f_n$ be the CPU frequency (cycle/s) of mobile device $n\in {\cal N}$.
Let $\rho_n f_n^{\zeta}$ denote the computational power of mobile device $n$, where $\rho_n$ is a constant that depends on the average switched capacitance and the average activity factor. The value of $\zeta (\zeta\geq 2)$ is a constant.
Then, the total energy consumption for local model training at mobile device $n$ is given by
\begin{equation}\label{e:2}
\text{E}_n^{\text{cmp}}=\log({1}/{\theta})\rho_nB_nD_nf_n^{\zeta-1}.
\end{equation}
Correspondingly, the total local computational latency for getting the optimal local parameters is expressed as $\text{T}_n^{\text{cmp}}=\log(1/\theta)(D_nB_n)/f_n, \forall n\in {\cal N}$.

As mentioned in Section II-B and II-C,  we can therefore conclude that the latency and energy consumption of one global iteration for mobile device $n$  are respectively defined as
\begin{equation}\label{eq:11}
\begin{array}{l}
\text{T}_{n}=\text{T}_{n}^{\text{cmp}}+\text{T}_{n}^{\text{up }}, \\
\text{E}_{n}=\text{E}_{n}^{\text{cmp}}+\text{E}_{n}^{ \text{up }}.
\end{array}
\end{equation}

\subsection{ Problem Formulation}
The proposed scheduling algorithm, which allows to determine 3M parameters $\{p_n^*, f_n^*, \theta, n\in{\cal N}\}$, pursuits to minimize mobile devices'  maximum latency and energy consumption.
To that end,  we employ the weighted sum method to deal with the tradeoff multiobjective using the tradeoff parameters $\lambda_n^{t}$ and $\lambda_n^{t}$ with $0\leq \lambda_n^{e}, \lambda_n^{t}\leq 1, \lambda_n^{e}+\lambda_n^{t}=1$, which describes the priority of each objective function of each device as follows:
\begin{equation}\label{e:3}
G_n({\boldsymbol p}, {\boldsymbol f}, \theta):\triangleq\lambda_n^{e}\text{E}_n+\lambda_n^{t}\text{T}_n,
\end{equation}
where ${\boldsymbol p}=[p_1, p_2, \ldots, p_N]^T$ and ${\boldsymbol f}=[f_1, f_2, \ldots, f_N]^T$ denote the transmit power profile and CPU frequency of mobile devices, respectively.
According to the aforementioned analysis, the upper bound of global iterations is normalized to $1/(1-\theta)$, and thus the total cost of mobile device $n$  in FL is calculated as
 \begin{equation}\label{e:4}
{\cal G}_n({\boldsymbol p}, {\boldsymbol f}, \theta):\triangleq\frac{1}{1-\theta}(\lambda_n^{e}\text{E}_n+\lambda_n^{t}\text{T}_n).
 \end{equation}
The balance coefficients $\lambda_n^{e}$ and $\lambda_n^{t}$ are determined by mobile devices' diverse demands and devices' resource conditions (e.g., the battery states and computing hardware configuration).
If a mobile device is with low battery, the device will schedule multi-resource to care more about energy consumption.
Likewise, the device will care more about latency in an action decision, when a device is running some applications that is sensitive to the latency (e.g., autonomous driving).
On one extreme, if $\lambda_n^{t}=1, \lambda_n^{e}=0$, then the mobile device is wholly latency-oriented; on the other extreme, if $\lambda_n^{t}=0, \lambda_n^{e}=1$, then the device is wholly energy consumption oriented.
Therefore, the optimal multi-resource allocation would like to solve the following optimization problem:
\begin{align}
 &\min _{\theta, \boldsymbol{f}, \boldsymbol{p}} \max_{n} {\cal G}_{n} \label{eq:12}\\
\text{s.t.~}  &\text{T}_{n}^{\text{cmp}}+\text{T}_{n}^{\text{up}} \leq \text{T}_{n}^{\max }, \forall n\in{\cal N}, \tag{\ref{eq:12}a}\\
 &\text{E}_{n}^{\text{up}} \leq \text{E}_{n, \max }^{\text{up}}, \forall n\in{\cal N},\tag{\ref{eq:12}b} \\
  &0 \leq f_{n} \leq f_{n}^{\max },  \forall n\in{\cal N},\tag{\ref{eq:12}c}\\
 &0 \leq \theta \leq 1,\tag{\ref{eq:12}d}
\end{align}
where $\text{E}_{n,\max }^{\text{up}}$ and $f_{n}^{\max }$ are the maximum transmission energy consumption and CPU frequency of device $n$.
The optimization problem (\ref{eq:12}) is non-convex, and is made particularly challenging by the constraint (\ref{eq:12}a) and several products of two functions in the objective function.
In the sequel, we present one computationally efficient approach to tackle (\ref{eq:12}).

\section{Scheduling Algorithm Design }
Solving the optimization problem (\ref{eq:12}) is challenging mainly due to several products in the objective function.
A tractable approach is to employ the alternating optimization technique to separately and iteratively solve for ${\boldsymbol p}, {\boldsymbol f}$, and local accuracy $\theta.$
We first solve for ${\boldsymbol p}, {\boldsymbol f}$ given $\theta$, and then obtain the optimal local accuracy $\theta$ when ${\boldsymbol p}, {\boldsymbol f}$ is fixed.
In the rest of this section, the optimization with respect to ${\boldsymbol p}, {\boldsymbol f}$ for fixed $\theta$, and with respect to $\theta$ for fixed ${\boldsymbol p}, {\boldsymbol f}$ will be treated separatively.
\subsection{Optimization for Block $\{\boldsymbol p, \boldsymbol f\}$}
For a fixed local accuracy $\theta$, the design problem (\ref{eq:12}) becomes the following convex problem
\begin{align}
&\min_{{\boldsymbol{f}},{\boldsymbol{p} } } ~ {\max_{n}{{\cal G}_{n}}} \label{eq:13}\\
\text{s.t.~} &\log \left ({\frac{1}{\theta } }\right )\frac{ D_{n}  B_{n}}{f_{n}} + \frac{C_{n}}{\sigma_{n}\sqrt[]{p_{n}} } \le \text{T}_{n}^{\max}, \tag{\ref{eq:13}a}\\
&p_{n} \le \left ( {\text{E}_{n,\max}^{\text{up}}\sigma_{n}}/{C_{n}}  \right )^{2}, \tag{\ref{eq:13}b}\\
& 0 \leq f_{n} \leq f_{n}^{\max }.\tag{\ref{eq:13}c}
\end{align}
To proceed further, by introducing an auxiliary variable $\xi$, the joint transmit power allocation and the corresponding devices' CPU scheduling problem (\ref{eq:13}) can thus be equivalent to
\begin{equation}\label{eq:14}
\begin{aligned}
&\min _{\xi,\boldsymbol{f},\boldsymbol{p}} \xi \\
\text{s.t.\quad}& \frac{1}{1-\theta}\left[\lambda_n^{t}\left(\log \left(\frac{1}{\theta}\right) \frac{D_{n} B_{n}}{f_{n}}+\frac{C_{n}}{\xi_{n} \sqrt{p_{n}}}\right)+\right. \\
&\left. \lambda_n^{e}\left(\log \left(\frac{1}{\theta}\right) \rho_{n}D_{n} B_{n} f_{n}^{\zeta-1}+\frac{C_{n} \sqrt{p_{n}}}{\sigma_{n}}\right)\right] \leq \xi,\\
&~(\ref{eq:13}\text{a}), (\ref{eq:13}\text{b}), \text{~and~} (\ref{eq:13}\text{c}).
\end{aligned}
\end{equation}

The problem is convex minimization problem, and hence the duality gap between (\ref{eq:14}) and its duality problem is zero. This means that the optimal solution of (\ref{eq:14}) can be obtained by applying the Lagrangian duality theory.
The Lagrangian function of (\ref{eq:14}) can be written as
\begin{align}\label{eq:15}
{\cal F}\left (\boldsymbol \lambda ,\boldsymbol \beta ,\boldsymbol \mu ,\boldsymbol \phi  \right ) = \min \nolimits_{\xi ,\boldsymbol f,\boldsymbol p}\mathcal{L}\left ( \xi ,\boldsymbol f,\boldsymbol p,\boldsymbol \lambda ,\boldsymbol \beta ,\boldsymbol \mu, \boldsymbol \phi \right),
\end{align}
where $\boldsymbol \lambda = \left \{ \lambda _{1},\lambda _{2},\cdots,\lambda _{N}\right \}\ge 0, \boldsymbol \beta = \left \{ \beta _{1},\beta _{2},\cdots,\beta _{N}\right \}\ge 0,\boldsymbol \mu = \left \{ \mu _{1},\mu _{2},\cdots,\mu _{N}\right \}\ge 0,\boldsymbol \phi = \left \{ \phi _{1},\phi _{2},\cdots,\phi _{N}\right \}\ge 0$ are the Lagrangian vector multipliers for the constraints in problem (\ref{eq:14}).

For the $n\text{-th}$ entry of Lagrangian vector multiplier $\pmb\mu$ of the CPU budget constraint (\ref{eq:13}c), define $\Xi=[0,-(Y_{1}^{{1}/{\zeta}}+
Y_{2}^{{1}/{\zeta}})] \cap(0, \varpi)$ is one possible subset of $\mu_n$, where $Y_{1}=\mu_{n}^{3}+6/(1-\theta) \lambda_n^{e} \phi_{n} \rho_{n} \log \left(\frac{1}{\theta}\right)D_{n} B_{n} \frac{-\mathcal{B}+\sqrt{\mathcal{B}^{2}-4 \mathcal{A C}}}{2}$,
$Y_{2}=\mu_{n}^{3}+6/(1-\theta) \lambda_n^{e} \phi_{n} \rho_{n} \log \left(\frac{1}{\theta}\right)D_{n} B_{n} \frac{-\mathcal{B}-\sqrt{\mathcal{B}^{2}-4 \mathcal{A C}}}{2},$
$\mathcal{C}=3 \mu_{n} \log \left(\frac{1}{\theta}\right)D_{n} B_{n}\left(\lambda_{n}+\lambda_n ^{t}/(1-\theta)\phi _{n}\right),$ and
$\varpi=3 \log \left(\frac{1}{\theta}\right)D_{n} B_{n}\left[\frac{1}{(1-\theta)^2}\left(\lambda_n^{e}\right)^{2} \phi_{n}^{2} \rho_{n}^{2}\left(\lambda_{n}+\lambda_n^{t}/(1-\theta) \phi_{n}\right)\right]^{\frac{1}{\zeta}}$.
Hence, drawing on dual theory techniques, we can obtain
\begin{equation}\label{eq:16}
f_{n}^{*}=\begin{cases}
\frac{-\mu_{n}-(Y_{1}^{\frac{1}{\zeta}}+Y_{2}^{{1}/{\zeta}})}{6 /(1-\theta) \lambda_n^{e} \phi_{n} \rho_{n} \log ({1}/{\theta})D_{n} B_{n}},& \text{if~} \mu_{n}\in\Xi, \\
\frac{\frac{18}{1-\theta}\lambda_n ^{e}\phi _{n}\rho_{n}\log^{2} \left (\frac{1}{\theta } \right ) D_{n}^{2}B_{n}\left (\lambda _{n} + \frac{\lambda_n ^{t}}{1-\theta}\phi _{n}\right )}{4 \mu_{n}^{2}},&\text{if~} \mu_{n}=\varpi, \\
\frac{\mu_{n}\left(\cos \frac{\arccos \mathcal{Q}}{\zeta}+\sqrt{\zeta} \sin \frac{\arccos \mathcal{Q}}{\zeta}-1\right)}{6 \mathcal{I}(\theta) \lambda^{e} \phi_{n} k_{n} \log \left(\frac{1}{\theta}\right)\left|D_{n}\right| q_{n}},&\text{if~} \mu_{n}>\varpi, \\
\text {no solution},&\text{otherwise,}
\end{cases}
\end{equation}
where $\mathcal{Q}=[\mu_{n} \mathcal{A}-\zeta/(1-\theta)\lambda_n^{e} \phi_{n} \rho_{n} \log \left(\frac{1}{\theta}\right ) D_{n} B_{n} \mathcal{B}]/{\mu_{n}^{\zeta}}$.

Likewise, the optimal solution of $\boldsymbol p$ is given by
\begin{equation}\label{eq:17}
p_{n}^{*}=\begin{cases}
\frac{\bar{\mathcal{A}}-\left(\sqrt[\zeta]{\bar{Y}_{1}}+\sqrt[\zeta]{\bar{Y}_{2}}\right)}{\zeta^2 \beta_{n}^2},&\text{if~}\frac{\phi_n\lambda_n^e}{1-\theta}\in\overline{\Xi}, \\
\frac{\phi_n^2\lambda_n^{e2}(\frac{\zeta+1}{\zeta} \sigma_{n}^{\zeta+1}-C_{n}^{\zeta+1})^2}{4(1-\theta)^2\beta _{n}^2C_{n}^{2\zeta} \sigma_{n}^2},& \text{if~}\frac{\phi_n\lambda_n^e}{1-\theta}= \widetilde\varpi,\\
\frac{\bar{\mathcal{A}}\left ( \cos {\frac{\arccos {\bar{\mathcal{Q}}}}{\zeta}+\sqrt[]{\zeta}\sin {\frac{\arccos {\bar{\mathcal{Q}}}}{\zeta}}-1 } \right )^2}{\zeta^2\beta_{n}^2} ,&\text{if~}\frac{\phi_n\lambda_n^e}{1-\theta}> \widetilde\varpi, \\
\text{no solution}, &\text{otherwise,}
\end{cases}
\end{equation}
where $\overline{\Xi}=[0,-{2 \sigma_{n}}/{C_{n}}(\sqrt[\zeta]{\bar{Y}_{1}}
+\sqrt[\zeta]{\bar{Y}_{2}})] \cap(0, \widetilde\varpi)$ denotes the possible set of $\frac{\phi_n\lambda_n^e}{1-\theta}$ and $\widetilde\varpi=\zeta \sqrt[\zeta]{\beta_{n}^{2} \frac{\sigma_{n}^{2}}{C_{n}^{2}}(\lambda_{n}+\frac{\lambda_n^{t}}{1-\theta}\phi_{n})}.$
Notably, $\bar{\mathcal{A}}= \frac{1}{4(1-\theta)^2}(\lambda_n^{e})^{2} \phi_{n}^{2} \frac{C_{n}^{2}}{\sigma_{n}^{2}},\bar{\mathcal{B}}=\frac{\zeta^2}{2} \beta_{n} \frac{C_{n}}{\sigma_{n}}\left(\lambda_{n}+\frac{\lambda_n^{t}}{1-\theta}\phi_{n}\right)$, $\bar{\mathcal{C}}=\frac{\zeta}{\zeta+1} 1/(1-\theta) \lambda_n^{e} \phi_{n} \frac{C_{n}^{2}}{\sigma_{n}^{2}}(\lambda_{n} +\lambda_n^{t}/(1-\theta)\phi_{n})$,
$\bar{Y}_{1}=\sqrt{\bar{\mathcal{A}}^{\zeta}}+\zeta \beta_{n} \frac{-\bar{\mathcal{B}}+\sqrt{\bar{\mathcal{B}}^{2}-(\zeta+1) \bar{\mathcal{A}} \bar{\mathcal{C}}}}{2},\bar{Y}_{2}=\sqrt{\bar{\mathcal{A}}^{\zeta}}+\zeta \beta_{n} \frac{-\bar{\mathcal{B}}-\sqrt{\bar{\mathcal{B}}^{2}-(\zeta+1) \bar{\mathcal{A}} \bar{\mathcal{C}}}}{2},$
and $\bar{\mathcal{Q}}=\frac{2 \sqrt{\bar{\mathcal{A}}^{\zeta}}-\zeta \beta_{n} \bar{\mathcal{B}}}{2 \sqrt{\bar{\mathcal{A}}^{\zeta}}}.$

The optimal solution structure of $f_n^*$ and $p_n^*$ have been obtained  given the local model accuracy $\theta$. According to the objective formula and the fist constriction in (\ref{eq:14}), the optimal solution of $\xi$ can be given by:
\begin{equation}\label{eq:18}
\begin{aligned}
\xi^{*}=&\max _{n} \frac{1}{1-\theta}\left[\lambda_n^{t}\left(\log \left(\frac{1}{\theta}\right) \frac{D_{n} B_{n}}{f_{n}^{*}}+\frac{C_{n}}{\sigma_{n} \sqrt{p_{n}^{*}}}\right)\right.\\
&\left.+\lambda_n^{e}\left(\log \left(\frac{1}{\theta}\right) k_{n}D_{n} B_{n}\left(f_{n}^{*}\right)^{2}+\frac{C_{n} \sqrt{p_{n}^{*}}}{\sigma_{n}}\right)\right].
\end{aligned}
\end{equation}

After obtaining the optimal ${\boldsymbol p}, {\boldsymbol f},$ and $\theta$, we update the Lagrangian vector multipliers of problem (\ref{eq:14}), i.e., ${\pmb \lambda}, {\pmb \beta}, {\pmb\mu},$ and $\pmb\phi$.
 It is well-known subgradient based method can be employed iteratively to find the optimal solutions for ${\pmb \lambda}, {\pmb \beta}, {\pmb\mu},$ and $\pmb\phi$.
Similar to the update of variable ${\boldsymbol p}, {\boldsymbol f},$ and $\theta$, the updates of ${\pmb \lambda}, {\pmb \beta}, {\pmb\mu},$ and $\pmb\phi$ are also separable.
Specifically, for $\lambda_n, \beta_n, \mu_n,$ and $\phi_n$, the pointwise update equations are given by
\begin{equation}\label{eq:19}
\begin{aligned}
\lambda_{n}(t+1)&=\max\{0,  ~\lambda_{n}(t)-i(t) \nabla \lambda_{n}(t)\}, \\
\beta_{n}(t+1)&=\max\{0,  ~\beta_{n}(t)-j(t) \nabla \beta_{n}(t)\}, \\
\mu_{n}(t+1)&=\max\{0,  ~\mu_{n}(t)-g(t) \nabla \mu_{n}(t)\}, \\
\phi_{n}(t+1)&=\max\{0,  ~\phi_{n}(t)-o(t) \nabla \phi_{n}(t)\},
\end{aligned}
\end{equation}
where $t$ is the iteration index and $i(t),j(t),g(t), o(t)\in(0, 1)$ are properly selected step size \cite{ramamonjison2014energy}.
All the mentioned sub-gradients are given by the following equation:
\begin{equation}\label{eq:20}
\begin{aligned}
\nabla\mu_{n}&=f_{n}^{*}-f_{n}^{\max },  \\
\nabla\beta_{n}&=p_{n}^{*}-({\text{E}_{n, \max }^{\text{up}} \sigma_{n}}/{C_{n}})^{2},  \\
\nabla\lambda_{n}&=\log (1/\theta) {D_{n}B_{n}}/{f_{n}^{*}}+(p_{n}^{*})^{-0.5}{C_{n}}/{\sigma_{n}},\\
\nabla\phi_{n}&=\left(\lambda_n^t\frac{B_nD_n}{f_n^*}+\lambda_n^e\rho_nB_nD_n(f_n^*)^2 \right)\frac{\log\left(\frac{1}{\theta}\right)}{1-\theta}\\
&~~+\left(\lambda_n^t\frac{C_n}{\sigma_n\sqrt{p_n^*}}+\lambda_n^e\frac{C_n\sqrt{p_n^*}}{\sigma_n}\right)
\frac{1}{1-\theta}-\xi^*.
\end{aligned}
\end{equation}

In the proposed alternating optimization algorithm, we solve ${\boldsymbol p}$ and $ {\boldsymbol f}$ by addressing (\ref{eq:14}) alternatively in an iterative manner, where the solution obtained in each iteration is used as the initial point of the next iteration.
The iterative optimization between $({\boldsymbol p}, {\boldsymbol f})$ and $(\pmb\lambda, \pmb\beta, \pmb\phi, \pmb\mu)$ is shown to converge to the optimal solution of problem (\ref{eq:12}) for given local accuracy.
The details of this procedure has been summarized in Algorithm 1.
\begin{algorithm}[!t]
\caption{Optimal ${\boldsymbol p}$ and ${\boldsymbol f}$ for (\ref{eq:14})}\label{alg:1}
\begin{algorithmic}[1]
\newcommand{\algorithmicInit}{\textbf{Initialization:}}
\newcommand{\algorithmicIter}{\textbf{Iteration:}}
\newcommand{\algorithmicOutput}{\textbf{Output:}}
\newcommand{\algorithmicBreak}{\textbf{break.}}
\REQUIRE\;

\algorithmicInit\;

{Set the initial value of dual variable to $\boldsymbol \lambda(0),\boldsymbol \beta(0) ,\boldsymbol \mu(0) ,\boldsymbol \phi(0)$, maximum number of iterations $\tau$ and the specific precision $\epsilon$.}

\algorithmicIter
\WHILE{$t \le \tau$}
\STATE{Substitute the dual variables $\boldsymbol \lambda(t),\boldsymbol \beta(t) ,\boldsymbol \mu(t)$, and $\boldsymbol \phi(t)$ into (\ref{eq:16}) and (\ref{eq:17}) to obtain $f_n(t)$ and $p_n(t)$, respectively;}
\STATE{Update new dual variables $\boldsymbol \lambda(t+1),\boldsymbol \beta(t+1) ,\boldsymbol \mu(t+1)$, and $\boldsymbol \phi(t+1)$ using (\ref{eq:19}), according to the new $f_n(t)$ and $p_n(t)$.}
\IF{$||\boldsymbol \lambda(t+1)-\boldsymbol \lambda(t)||<\epsilon,||\boldsymbol \beta(t+1)-\boldsymbol \beta(t)||<\epsilon,||\boldsymbol \mu(t+1)-\boldsymbol \mu(t)||<\epsilon$, and$||\boldsymbol \phi(t+1)-\boldsymbol \phi(t)||<\epsilon$}
\STATE{$f_n^*=f_n(t)$ and $p_n^*=p_n(t)$.}\;
\STATE{\algorithmicBreak}
\ELSE
\STATE{$t=t+1$}
\ENDIF
\ENDWHILE

\algorithmicOutput {~$\boldsymbol f^*=(f_1^*,\ldots,f_N^*)$ and $\boldsymbol p^*=(p_1^*,\ldots,p_N^*).$}
\end{algorithmic}
\end{algorithm}

\subsection{Optimization for $\theta$}
We now turn again our attention in problem (\ref{eq:13}) for the case where ${\boldsymbol p}$ and ${\boldsymbol f}$ are fixed and the objective is the optimization over $\theta.$
Particularly, we focus on solving:
\begin{align}
\min_\theta\max_{n}&\frac{1}{1-\theta}\left[\log\left(\frac{1}{\theta}\right)
\left(\frac{\lambda_n^tD_nB_n}{f_n}+\lambda_n^e\rho_nD_nB_nf_n^2\right)\right.\notag\\
&\left.+\lambda_n^t\frac{C_n}{\sigma_n\sqrt{p_n}}+\lambda_n^e
\frac{C_n\sqrt{p_n}}{\sigma_n}\right] \label{e:6}\\
\text{s.t.~}&\log\left(\frac{1}{\theta}\right)\frac{D_nB_n}{f_n}\leq
\text{T}_{n,\max}^{\text{up}}-\frac{C_n}{\sigma_n\sqrt{p_n}}. \tag{\ref{e:6}a}
\end{align}
It is can be that, for fixed ${\boldsymbol p}$ and ${\boldsymbol f}$, problem (\ref{e:6}) can be transformed into unconstrained problem by constructing penalty part from (\ref{e:6}a).
As a consequence, for $\theta\in[0, 1]$, problem (\ref{e:6}) can be written as
\begin{align}
\min_{\theta}\frac{1}{1-\theta}&\left[\log\left(\frac{1}{\theta}\right)
\left(\frac{\lambda_n^tD_nB_n}{f_n}+\lambda_n^e\rho_nD_nB_nf_n^2\right)\right.\notag\\
&\left.+\lambda_n^t\frac{C_n}{\sigma_n\sqrt{p_n}}+\lambda_n^e
\frac{C_n\sqrt{p_n}}{\sigma_n}\right]
+\frac{1}{\delta}\sum_{n=1}^N\max\left\{ 0,\right.\notag\\
&\left.\frac{D_nB_n}{f_n}\log\left(\frac{1}{\theta}\right)
-\frac{C_n}{\sigma_n\sqrt{p_n}}-\text{T}_{n,\max}^{\text{up}}\right\}. \label{e:7}
\end{align}

To solve (\ref{e:7}), we use a self-adaptive global best harmony search algorithm to solve this continuous optimization problem.
The main parameters of the algorithm are shown below:
\begin{itemize}
\item Harmony memory size (HMS): setting $\texttt{HMS}=5$;

\item Harmony memory consideration rate (HMCR): We assume that the $\texttt{HMCR}$ value is normally distributed in the tange of $[0.9,1]$ with mean $0.98$ and standard deviation $0.01$.

\item Pitch adjustment rate (PAR): The $\texttt{PAR}$ value is distributed in the range of $[0,1]$ with mean $0.9$ and standard deviation $0.05$.

\item Distance Bandwidth (BW): Let BW be updated with the following formula:
\begin{equation}\label{eq:23}
\hspace{-0.1cm}\texttt{BW}(t)=\begin{cases}
\texttt{BW}_{\max}-\frac{\texttt{BW}_{\max}-\texttt{BW}_{\min}}{\text{T}_{\max}} 2t,&\hspace{-0.25cm}\text{if~} t<\frac{\text{T}_{\max}}{2}, \\
\texttt{BW}_{\min},&\hspace{-0.25cm}\text{if~}t\ge\frac{\text{T}_{\max}}{2},
\end{cases}
\end{equation}
where $\texttt{BW}_{\min}=0.0005$ and $\texttt{BW}_{\max}=0.05$ are the minimum and maximum distance bandwidths, respectively.

\item Number of Improvisations: setting $\text{T}_{\max}=5000.$
\end{itemize}
The details of the proposed algorithm for (\ref{e:7}) are presented in Algorithm \ref{alg:2}.
\begin{algorithm}[!t]
\caption{  Harmony Search for Problem (\ref{e:7})}\label{alg:2}
\begin{algorithmic}[1]
\newcommand{\algorithmicInit}{\textbf{Initialization:}}
\newcommand{\algorithmicIter}{\textbf{Iteration:}}
\newcommand{\algorithmicOutput}{\textbf{Output:}}
\newcommand{\algorithmicBreak}{\textbf{break.}}
\REQUIRE\;

\algorithmicInit\;

{Set the parameters $\texttt{HMS}$ and $\text{T}_{\max}.$}

{Initialize the HM and calculate the objective function value of each harmony vector.}

{Set $t=1$ and $l_1,l_2,l_3 \in (0,1)$.}

\algorithmicIter
\WHILE{$t \le \text{T}_{\max}$}
\STATE{Generate $\texttt{HMCR}$ and $\texttt{PAR}$ using the mentioned normally distribution. And compute $\texttt{BW}(t)$ according to (\ref{eq:23}).}
\IF{$l_1<\text{T}_{\max}$}
\STATE{$\theta^{+}=\theta_h \pm l\times \texttt{BW}$, where $h\in \{1,2,\cdots,\texttt{HMS}\}$.}
\IF{$l_2<\texttt{PAR}$}
\STATE{$\theta^{+}=\theta^{\text{o}}$, where $\theta^{\text{o}}$ is the best harmony in the HM as evaluated by objective function, $l_1,l_2\in(0,1)$.}
\ENDIF
\ELSE
\STATE{$\theta^{+}=\theta_{\min}+l\times (\theta_{\max}-\theta_{\min})$, where $l\in(0,1)$}
\ENDIF
\IF{The objective value at $\theta^{+}$ is less than that at $\theta^{\text{w}}$, i.e., the worst harmony in the HM}
\STATE{Substitute $\theta^{\text{w}}$ in HM to $\theta^{+}$}
\ENDIF
\STATE{$t=t+1$}
\ENDWHILE

\algorithmicOutput {~The optimal local model accuracy $\theta^*$ in the HM as evaluated by objective function in (\ref{e:7}).}
\end{algorithmic}
\end{algorithm}

\begin{figure*}[!t]
\centering
	\subfigure[]{
		\begin{minipage}[t]{0.3\linewidth}
			\includegraphics[width=2in]{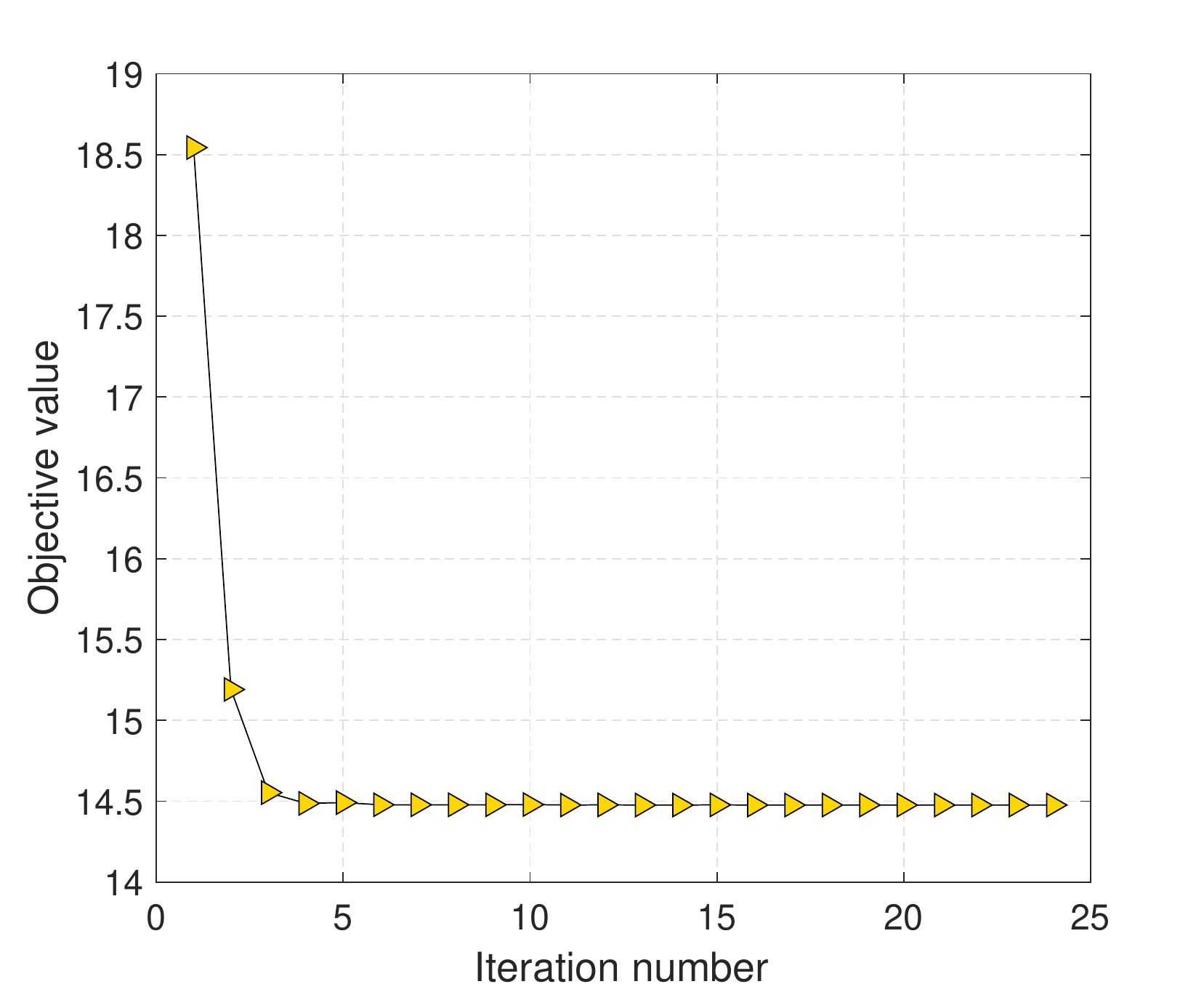}
		\end{minipage}
		\label{fig:2-1}
	}
	\subfigure[]{
		\begin{minipage}[t]{0.3\linewidth}
			\includegraphics[width=2in]{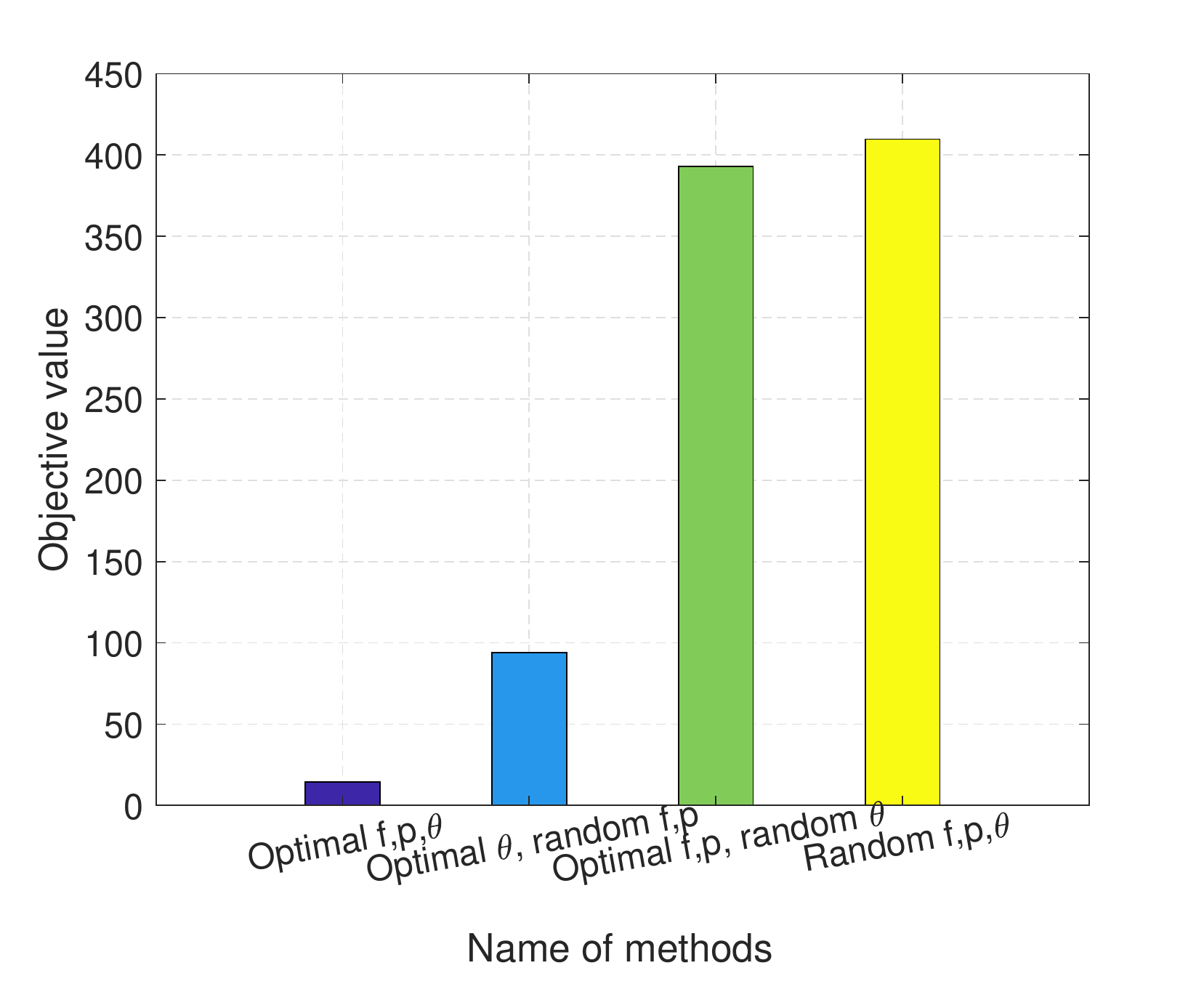}
		\end{minipage}
		\label{fig:2-2}
	}
	\subfigure[]{
		\begin{minipage}[t]{0.3\linewidth}
			\includegraphics[width=2in]{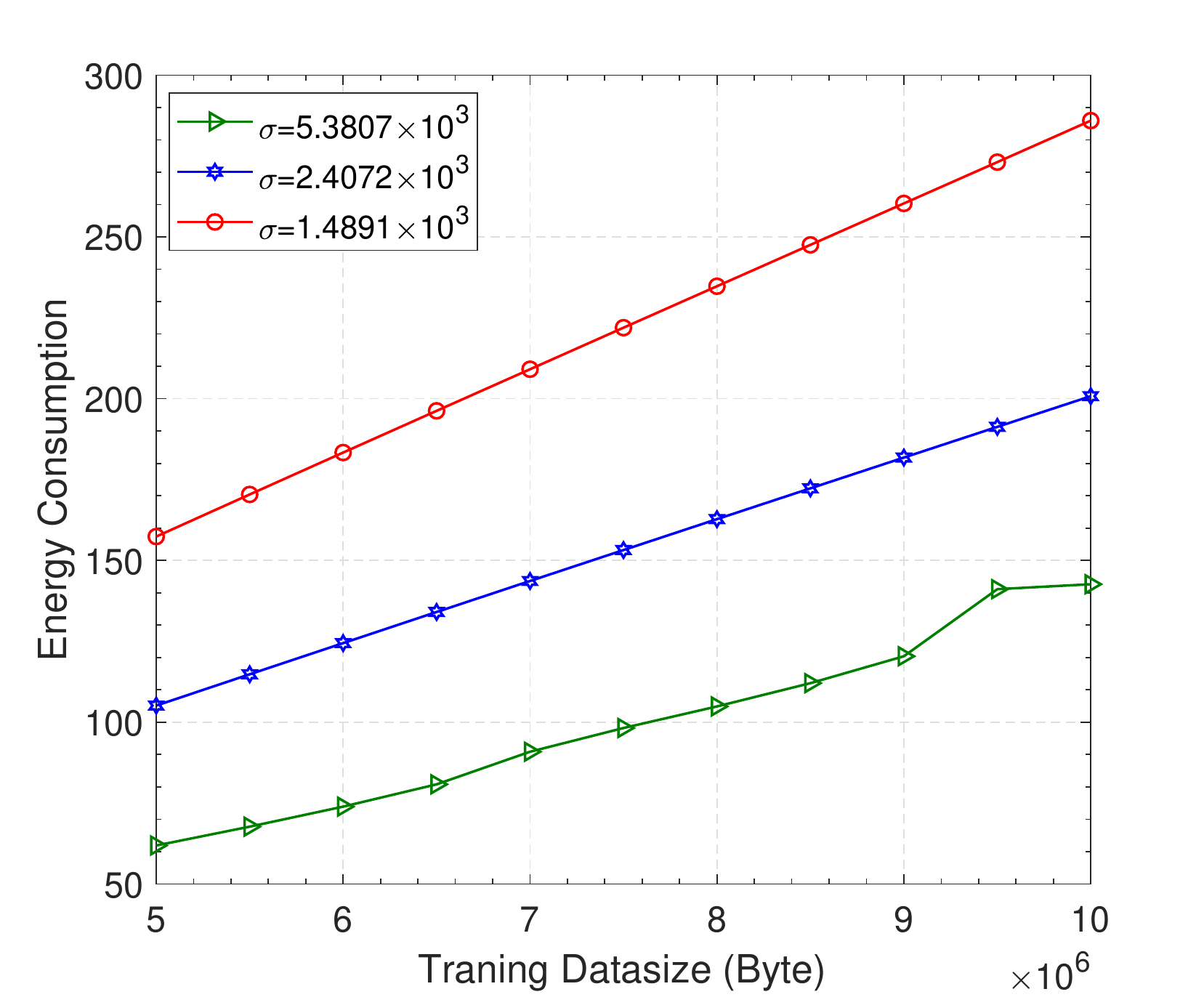}
		\end{minipage}
		\label{fig:2-3}
	}
	\caption{(a) Objective value, (b) algorithms comparison, and (c) energy consumption under different factors.}
	\label{fig:2}
\end{figure*}
\section{Simulation Results}
In this section, we conduct extensive simulations to assess the performance of our proposed multi-resource allocation algorithm for the on-device distributed FL system.
The convergence property and effectiveness of the proposed alternating optimization algorithm are verified.
The simulation environment settings are as follows.
We consider the scenario shown in Fig.\ref{fig:1}, where the wireless edge server has a coverage range of $150\text{~m}$ and the FL area is $100\text{~m},$ and the number of devices passing the edge server per unit of time follows a Poisson process.
Throughout the simulations, unless otherwise specified, we adopt the remaining parameters as follows.
Following \cite{gudmundson1991correlation}, the positive constant $A=90.2514, B=3,4998, C=1.0942,$ and for simplicity, each mobile device is assumed to have the same latency constraint and the maximum CPU frequency, that is, $\text{T}_n^{\max}=\text{T}^{\max}=4\text{~s}$ and $f_n^{\max}=f^{\max}=2.0\text{~GHz}.$
For the size of dataset ${\cal D}_n$ of mobile device $n\in{\cal N}$ is randomly assigned from the interval $[5, 10]\text{~MB}$ to account for the heterogeneous data of mobile devices.
Moreover, for each $ n\in{\cal N}$, the average CPU cycle is $B_n=40$ (cycles/byte), the size of model parameters $C_n=4.5\text{~KB},$ the effective switching capacity $\rho_n=10^{-24}$, and the energy budget is $\text{E}_{n, \max}^{\text{up}}=20\text{~J}.$
According to \cite{zhu2022intelligent,ma2022distributed}, the relationship between the power computation and CPU frequency is given as below, namely, $\rho_n=0.05$ and $\zeta=3.$
Additionally, the weights are related by $\lambda_n^{e}=1-\lambda_n^{t}, \forall n\in{\cal N},$ for simplicity, each mobile device is assumed to have the same scaler weight, $\lambda_n^{t}=\lambda^{t}=1/2$ and $\lambda_n^{e}=\lambda^{e}=1/2. $
For on-device distributed federated learning systems, we evaluate the performance of the proposed iteratively algorithm with the following three methods: (i) method  optimizing $\theta$ as well as  randomly selecting ${\boldsymbol f}$ and ${\boldsymbol p}$; (ii) method optimizing ${\boldsymbol p}$ and ${\boldsymbol f}$ as well as randomly selecting $\theta$; (iii) method randomly selecting ${\boldsymbol p}, {\boldsymbol f},$ and $\theta.$
Figure \ref{fig:2-1} illustrates the convergence of our proposed multi-resource allocation algorithm.
As seen in Fig. \ref{fig:2-1}, our proposed alternating algorithm can quickly converges within $5$ iterations.
Further performance comparison of different methods is shown in Fig. \ref{fig:2-2}.
Specifically, in Fig. \ref{fig:2-2}, for optimization problem (\ref{eq:14}), we assess the performance of weighted sum of latency and energy consumption under different methods.
From the results, as expected, the proposed alternatively algorithm outperforms the other remaining methods.
Compared with the proposed algorithm, the increase of the objective value for method (ii) is significantly, this can be inferred by the upper bound of global iterations $1/(1-\theta).$
Finally, we investigate the impact of the size of training datasets on the optimization algorithm, by adjusting the size of datasets from $5\times 10^6$ to $10\times 10^6$ KB.
As shown in Fig.\ref{fig:2-3}, energy consumption increases with the size growth of datasets. This is because devices require more energy and time to process large amounts of data.

\section{Conclusions}
In this paper, we studied the joint transmit power/computation resource allocation and local model accuracy optimization for on-device distributed federated learning systems.
Specifically, the behavior of the device in the network can be divided into a local computing phase and a data transmission phase. Our goal is to minimize the cost function by optimizing the CPU frequency, transmit power and local model accuracy while satisfying the latency and CPU constraints. We developed the iterative optimization algorithm based on the Lagrange multiplier and harmonic search method in order to obtain the global optimal solution. Simulation results validated the proposed algorithm's effectiveness.


\end{document}